\documentclass[11pt, twoside, usenatbib, usepslatex]{article}

\usepackage{asp2006}
\usepackage{epsf}
\usepackage{psfig}
\usepackage{graphicx}
\usepackage{natbib}
\usepackage{subfigure}
\usepackage{ctable}
\usepackage{multirow}
\begin{document}

\title{Are low-degree p mode frequencies predictable from one cycle to the next?}

\author{A.~M. Broomhall$^1$, W.~J. Chaplin$^1$,
Y. Elsworth$^1$, R. New$^2$, G.~A. Verner$^3$}

\affil{$^1$School of Physics and
Astronomy, University of Birmingham, Edgbaston, Birmingham B15 2TT\\
$^2$Faculty of Arts, Computing, Engineering and Sciences, Sheffield Hallam University, Sheffield S1 1WB\\
$^3$Astronomy Unit, School of Mathematical Sciences, Queen Mary, University of London, Mile End Road, London E1 4NS }    %%% Fill in author affiliations

\begin{abstract}
The Birmingham Solar-Oscillations Network (BiSON) has been
collecting data for over $30\,\rm yrs$ and so observations span
nearly three $11\,\rm yr$ solar activity cycles. This allows us to
address important questions concerning the solar cycle and its
effect on solar oscillations, such as: how consistent is the
acoustic behaviour from one cycle to the next? We have used the
p-mode frequencies observed in BiSON data from one solar activity
cycle (cycle 22) to predict the mode frequencies that were observed
in the next activity cycle (cycle 23). Some bias in the predicted
frequencies was observed when short $108\,\rm d$ time series were
used to make the predictions. We also found that the accuracy of the
predictions was dependent on which activity proxy was used to make
the predictions and on the length of the relevant time series.
\end{abstract}

\section{Introduction}\label{section[intro]}
The frequencies of solar oscillations vary with solar activity
\citep{Woodard1985} and the changes in frequency can, in principle,
be used to infer solar-cycle related structural changes in the Sun.
Methods have been developed to remove solar cycle frequency shifts
\citep[e.g.][]{Chaplin2004} to allow comparisons between the mode
frequencies obtained from non-contemporaneous data and to enable
accurate activity-independent structural inversions to be performed.

The Birmingham Solar-Oscillations Network (BiSON) has been recording
data for over $30\,\rm yrs$, and so observations span nearly three
solar cycles. We aim to take advantage of this to answer the
following question: can the mode frequencies observed in cycle 22 be
used to predict the mode frequencies observed in cycle 23? In other
words, is it possible to produce a model that can be used to predict
the mode frequencies observed in cycle 23?

Each $11\,\rm yr$ solar cycle is different from the previous one,
for example, cycle 23 has been described as magnetically simpler
than recent previous cycles \citep{Toma2004}. Several different
activity proxies can be used to measure the level of solar activity
and, as each proxy responds to different physical processes, the
amplitude and shape of each activity cycle depends upon which
activity proxy is considered. With this in mind we aim to determine
whether the acoustic behaviour of the Sun is predictable from one
cycle to the next.

We begin this paper by describing how we used the mode frequencies,
observed in cycle 22, to predict the mode frequencies observed in
cycle 23 (Section 2). Then, in Section 3, we compare these predicted
frequencies with the actual frequencies observed in cycle 23.
Section 4 comprises a discussion of the results.

\section{Predicting frequency shifts}\label{section[method]}
An $11\,\rm yr$ BiSON time series is available, which covers the
whole of cycle 22. To allow solar cycle frequency shifts to be
examined this data set was split into contiguous $108\,\rm d$ time
series. A standard likelihood maximization method was used to fit
the power spectra of these time series \citep[e.g.][]{Chaplin1999},
enabling the observed mode frequencies to be determined. The
weighted mean activity level, $a$, observed during each $108\,\rm d$
time period was also calculated for four activity proxies, namely
the $10.7\,\rm cm$ flux $(\rm F_{10.7})$, the HeI equivalent width
(HeI EW), the Kitt Peak Magnetic Index (KPMI) and the International
Sunspot Number (ISN).

A minimum activity reference set was determined by averaging the
frequencies from five $108\,\rm d$ time series that were observed
during the minimum activity epoch at the boundary between cycle 22
and cycle 23. We define the solar cycle frequency shifts as the
difference between the frequencies given in the minimum activity
reference set and the frequencies of the corresponding modes
observed at different epochs and, consequently, different levels of
activity. The size of a frequency shift has a well-known dependency
on frequency and mode inertia and these dependencies were removed in
the manner described in \citet{Chaplin2004}. We have, therefore,
determined the mean frequency shift at a given activity level. We
now describe how these shifts were used to predict the mode
frequencies observed in cycle 23.

\begin{figure}
  \centering
  \includegraphics[clip, width=4.1cm]{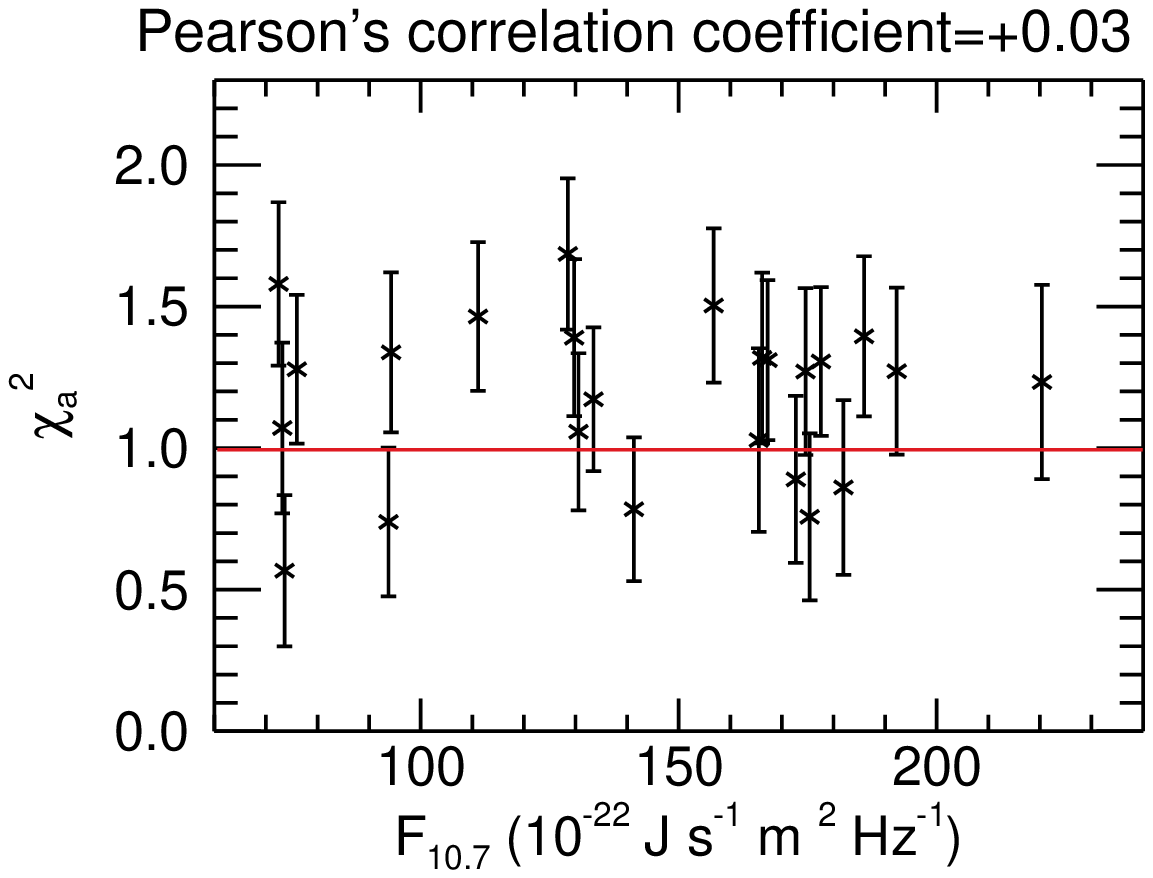}
  \includegraphics[clip, width=4.1cm]{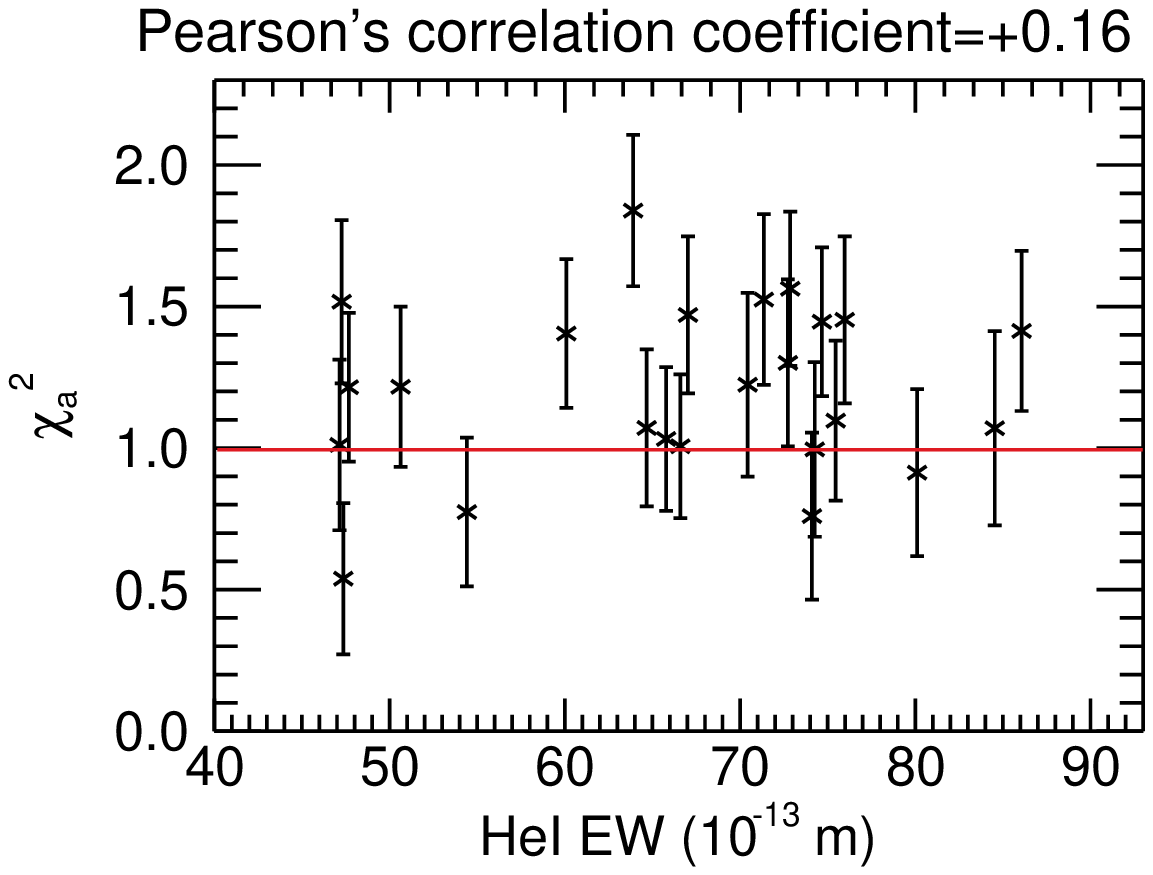}\\
  \includegraphics[clip, width=4.1cm]{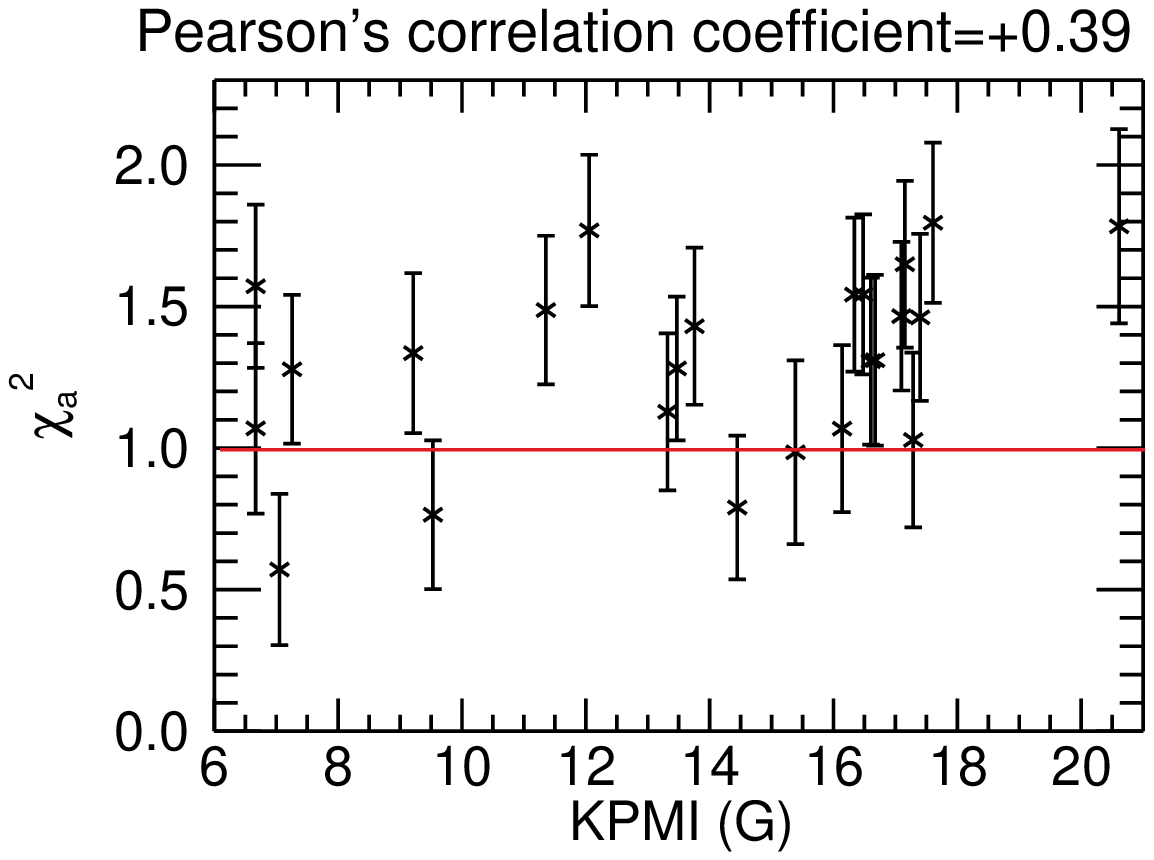}
  \includegraphics[clip, width=4.1cm]{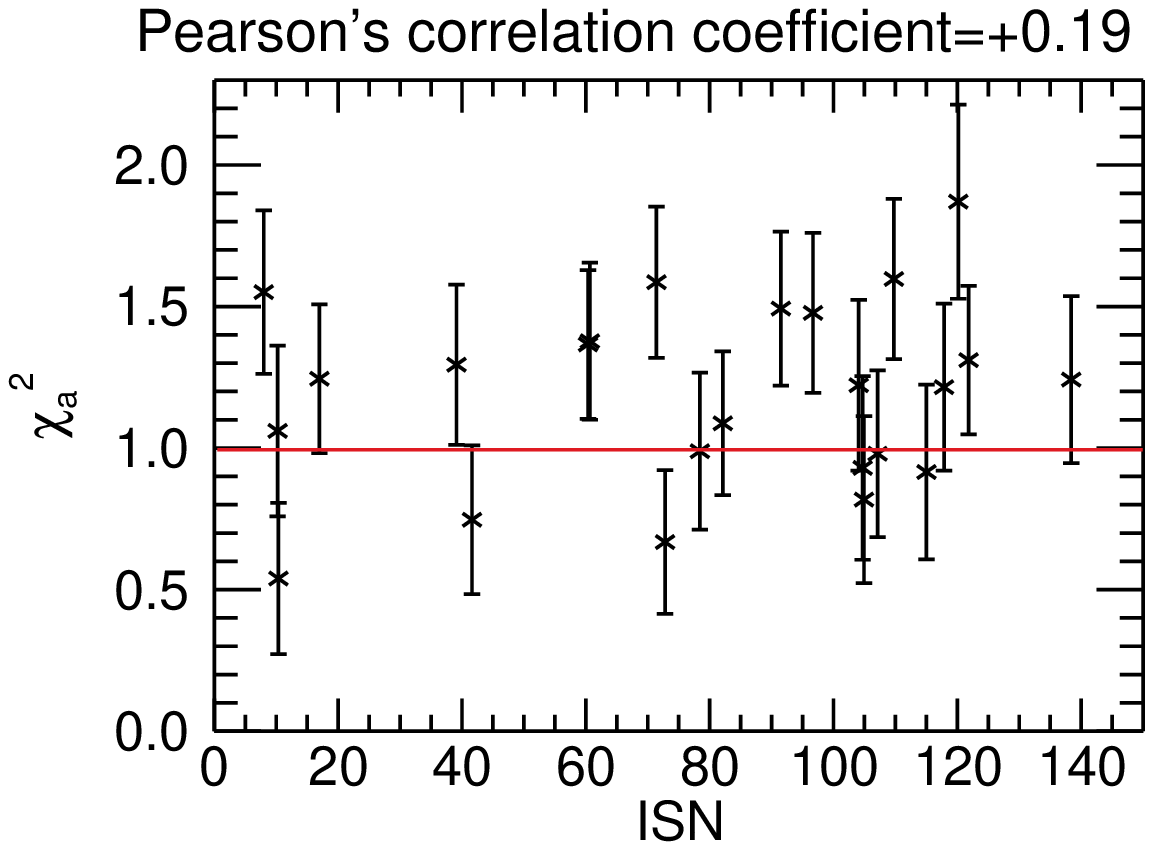}\\
  \vspace{-0.3cm}
  \caption{The $\chi_a^2$ observed at different activity levels when four
  different activity proxies were used to predict the frequency
  shifts.}
  \label{figure[chi a]}
\end{figure}

For low-degree p modes, the relationship between activity and mean
frequency shift is approximately linear. A linear least squares fit
was performed to determine the coefficients of the relationship
between the mean frequency shifts observed during cycle 22 and the
corresponding mean activity observed by each of the four activity
proxies. These linear relationships were then used to predict the
size of the mean frequency shift at a given activity level, which
corresponds to a particular time period from cycle 23.

Let $\delta\nu_1(a)$ be the mean frequency shift predicted by the
linear relationship for a given activity level, $a$. To determine
the frequency shift for a particular mode, $\delta\nu_{n,l}(a)$,
with degree, $l$, and radial order, $n$, it was necessary to scale
$\delta\nu_1(a)$ with mode frequency and inertia. This was done
using equation 4 of \citet{Chaplin2004}. The resulting frequency
shift, $\delta\nu_{n,l}(a)$, was added to the corresponding mode
frequency stored in the minimum activity reference set to give a
predicted frequency, and its associated errors, for a given epoch.
The predicted frequencies were compared to the frequencies observed
in BiSON data during the same time span and we now go on to describe
the results of this comparison.

\begin{table}[t]
  \centering
  \caption{The weighted mean $\chi_a^2$ for the different activity proxies}
  \vspace{0.1cm}
  \begin{tabular}{ccc}
    \hline
   \textbf{Proxy} & \textbf{Weighted mean $\chi_a^2$} & \textbf{Distance from $\chi^2=1$}\\
    \hline
    $\rm F_{10.7}$ & $1.18\pm0.06$ & $3.1\sigma$\\

    HeI EW & $1.20\pm0.06$ & $3.5\sigma$\\

    KPMI & $1.30\pm0.06$ & $5.2\sigma$\\

    ISN & $1.18\pm0.06$ & $3.1\sigma$\\
    \hline
  \end{tabular}
  \label{table[total chi]}
\end{table}

\section{Analysis of results}\label{section[results]}
To determine how accurate the prediction method is we have compared
the predicted frequency of a mode at an activity level, $a$, with
the frequency observed in BiSON data at the same activity level. The
observed frequencies were determined by fitting the power spectrum
of the corresponding $108\,\rm d$ time series using the standard
likelihood maximization method \citep{Chaplin1999}.

The reduced $\chi^2$ between the predicted and the observed
frequencies at a particular activity level was determined. Let
$\chi_a^2$ be the reduced $\chi^2$ for all modes for which accurate
fits to the data were obtained in a $108\,\rm d$ time series,
observed during cycle 23. Errors on the reduced $\chi^2$ were taken
to be $\sqrt{2/(N-1)}$, where $N$ is the number of data points used
to calculate the reduced $\chi^2$. Figure \ref{figure[chi a]} shows
the variation of $\chi_a^2$ with activity, where each panel shows
the results predicted by a different activity proxy. The vast
majority of the observed $\chi_a^2$ are within $3\sigma$ of unity,
which is the expected value if the model is accurate. However, for
all activity proxies, the observed $\chi_a^2$ are predominately
greater than unity, which implies there is some bias in the
predicted results.

\begin{figure}
\centering
\begin{tabular}{cc}
  \includegraphics[clip, width=4.0cm]{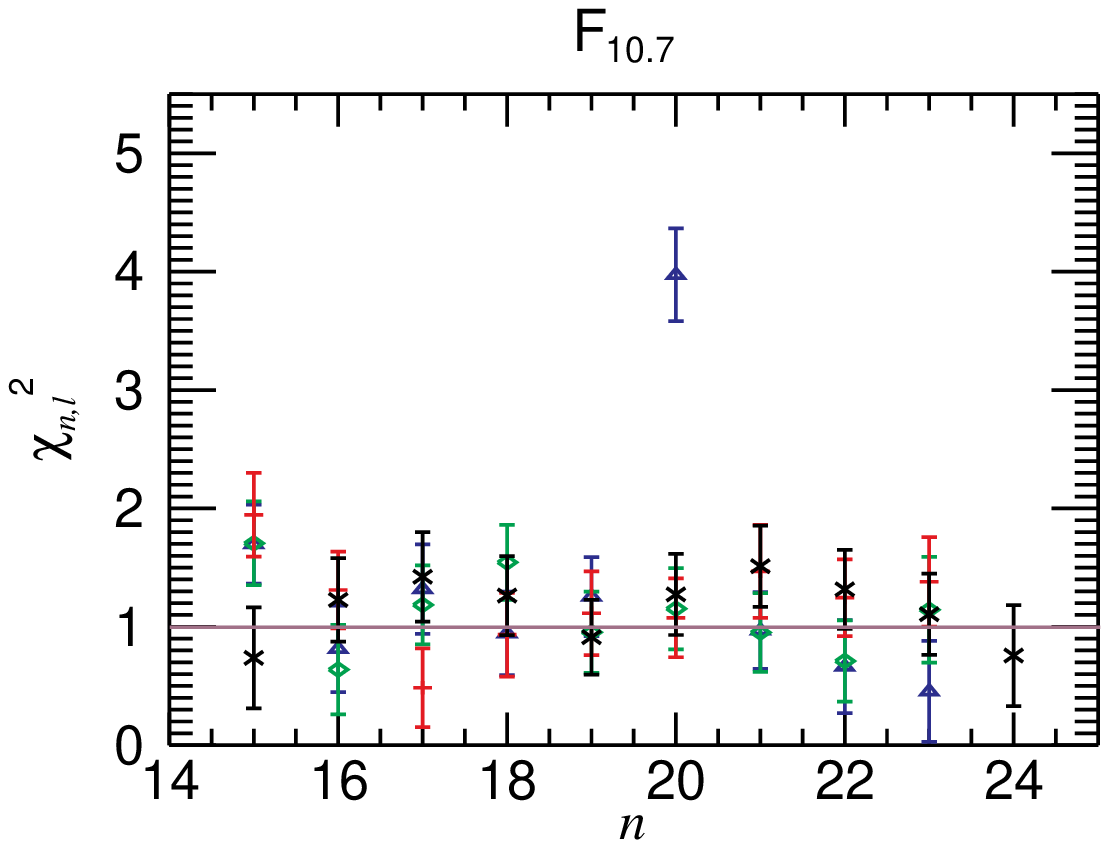}
  \includegraphics[clip, width=4.0cm]{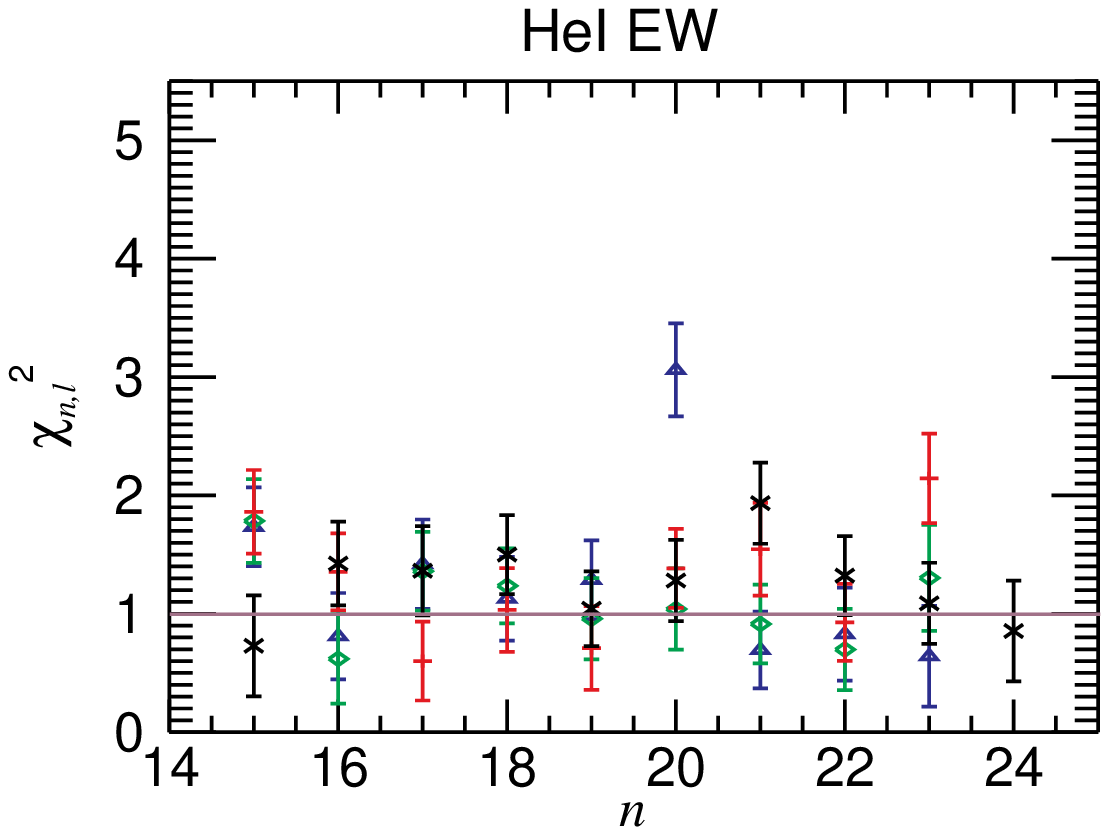} &  \multirow{2}{*}{\includegraphics[clip,
  width=1.2cm]{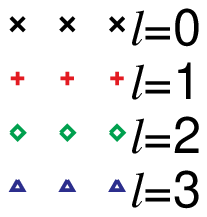}}\\
   \includegraphics[clip, width=4.0cm]{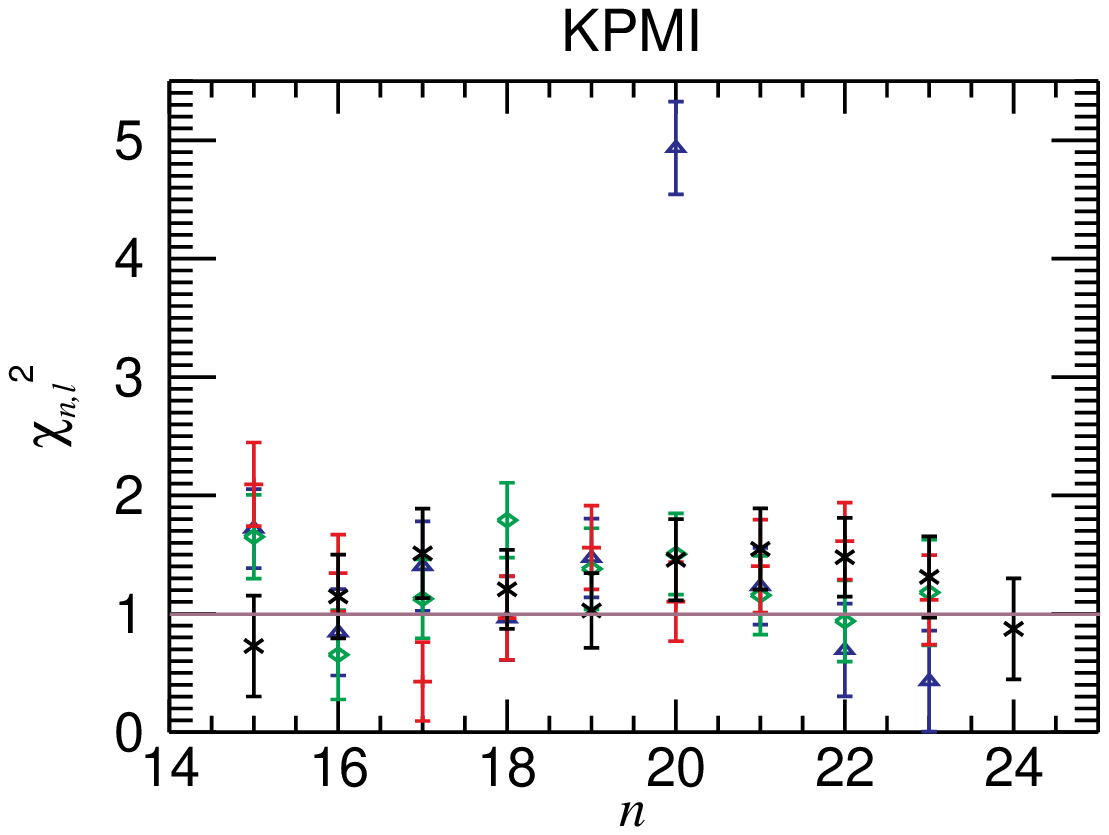}
  \includegraphics[clip, width=4.0cm]{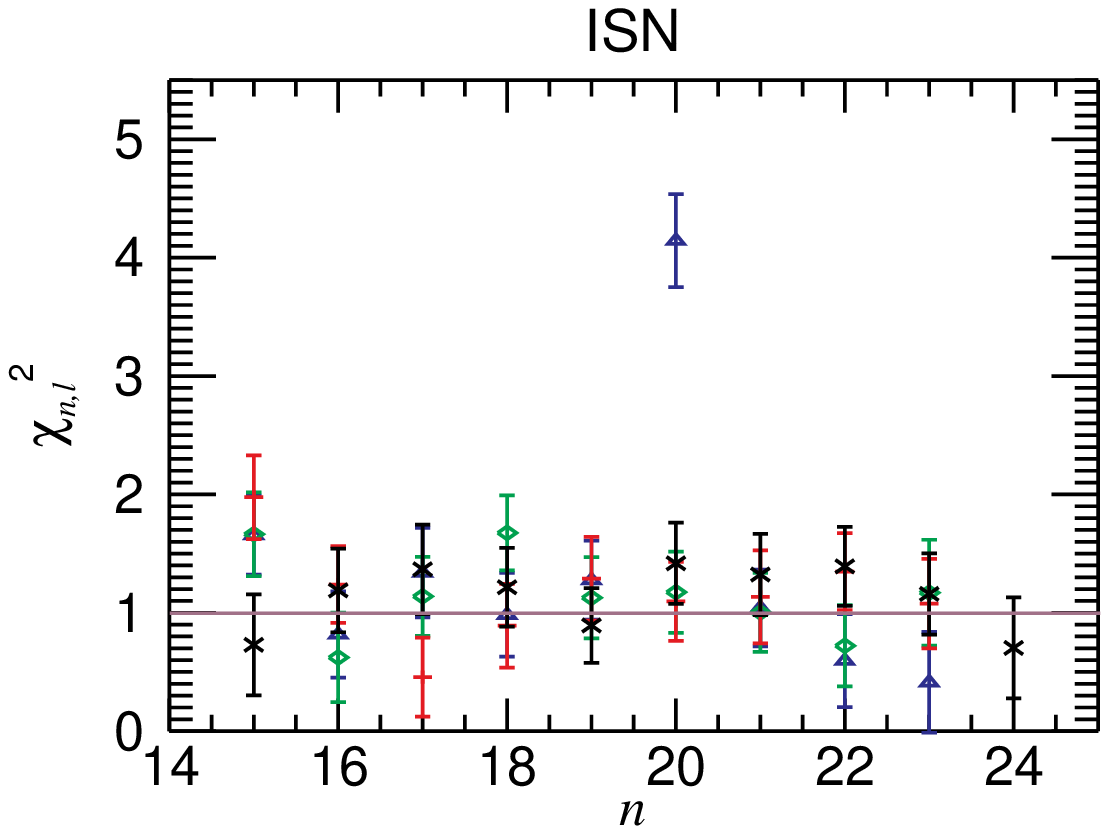} & \\
\end{tabular}
\vspace{-0.3cm}
   \caption{The $\chi_{n,l}^2$ observed at different radial orders, $n$, when four
  different activity proxies were used to predict the frequencies.}\label{figure[radial chi]}
\end{figure}

To determine whether there is any correlation between the level of
activity and the accuracy of the predicted frequencies the Pearson's
correlation coefficient between activity level and $\chi_a^2$ was
calculated (see the values above the panels in Figure
\ref{figure[chi a]}). None of the correlation coefficients are
significant at a 95\% confidence level. However, the correlation
coefficient for the $10.7\,\rm cm$ flux result is smaller than for
the other activity proxies. The results imply that the accuracy of
the predictions made using the HeI EW, the ISN and the KPMI are
marginally dependent on activity. We propose that this is true for
the following reason. The HeI EW is mainly sensitive to the weak
magnetic field and the ISN and KPMI are most sensitive to the strong
magnetic field, while the $10.7\,\rm cm$ flux shows good sensitivity
to both the strong and the weak magnetic flux. Therefore, we believe
that to make accurate predictions at all activity levels it is
necessary for the activity proxy to be responsive to both the strong
and weak flux.

As a measure of the overall accuracy of the predictions the weighted
mean $\chi_a^2$ was determined, and the results are shown in Table
\ref{table[total chi]}. All of the weighted mean $\chi_a^2$ are more
than $3\sigma$ from unity, indicating there is some bias in the
predicted frequencies. The weighted mean $\chi_a^2$ is noticeably
larger for the KPMI than when the other proxies are used, which
highlights the inadequacies of this proxy as a predictive tool.

\begin{figure}[b]
    \centering
    {\includegraphics[clip,
    width=3.9cm]{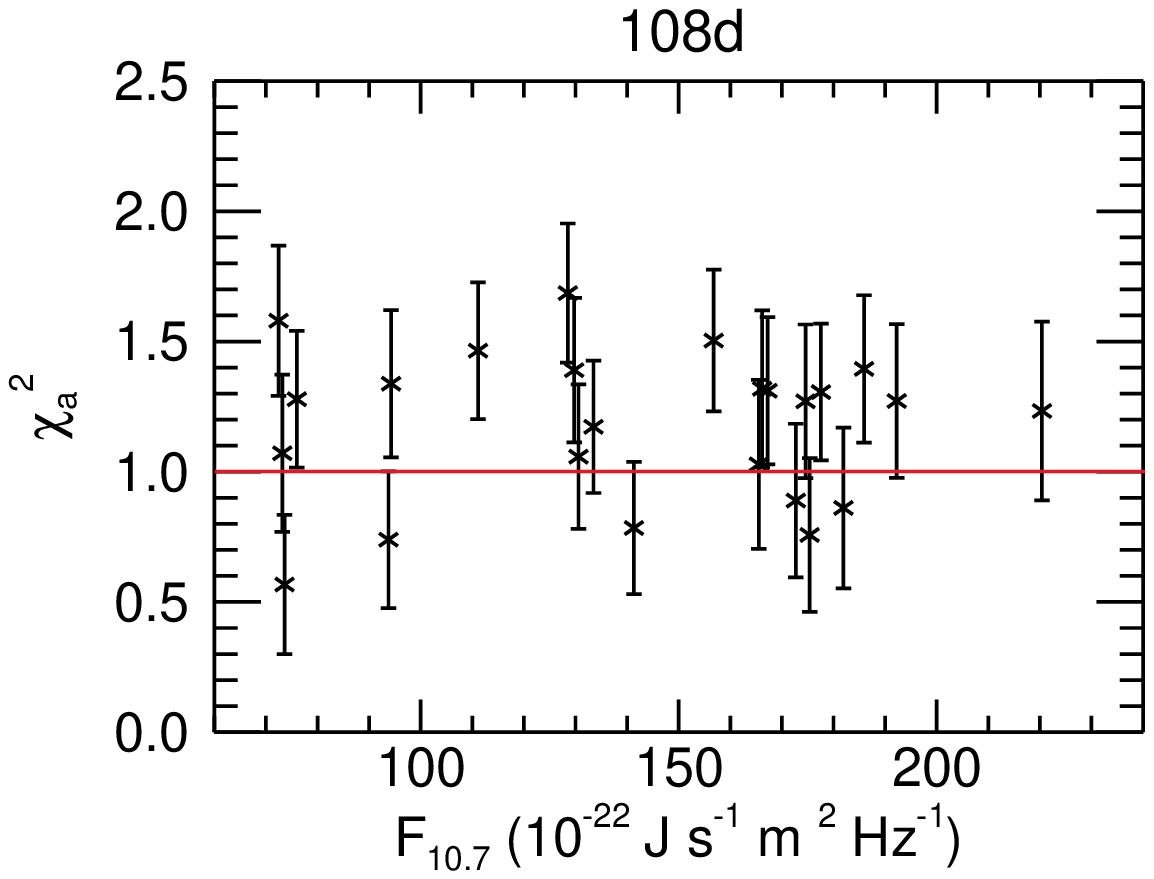}}
    {\includegraphics[clip,
    width=3.9cm]{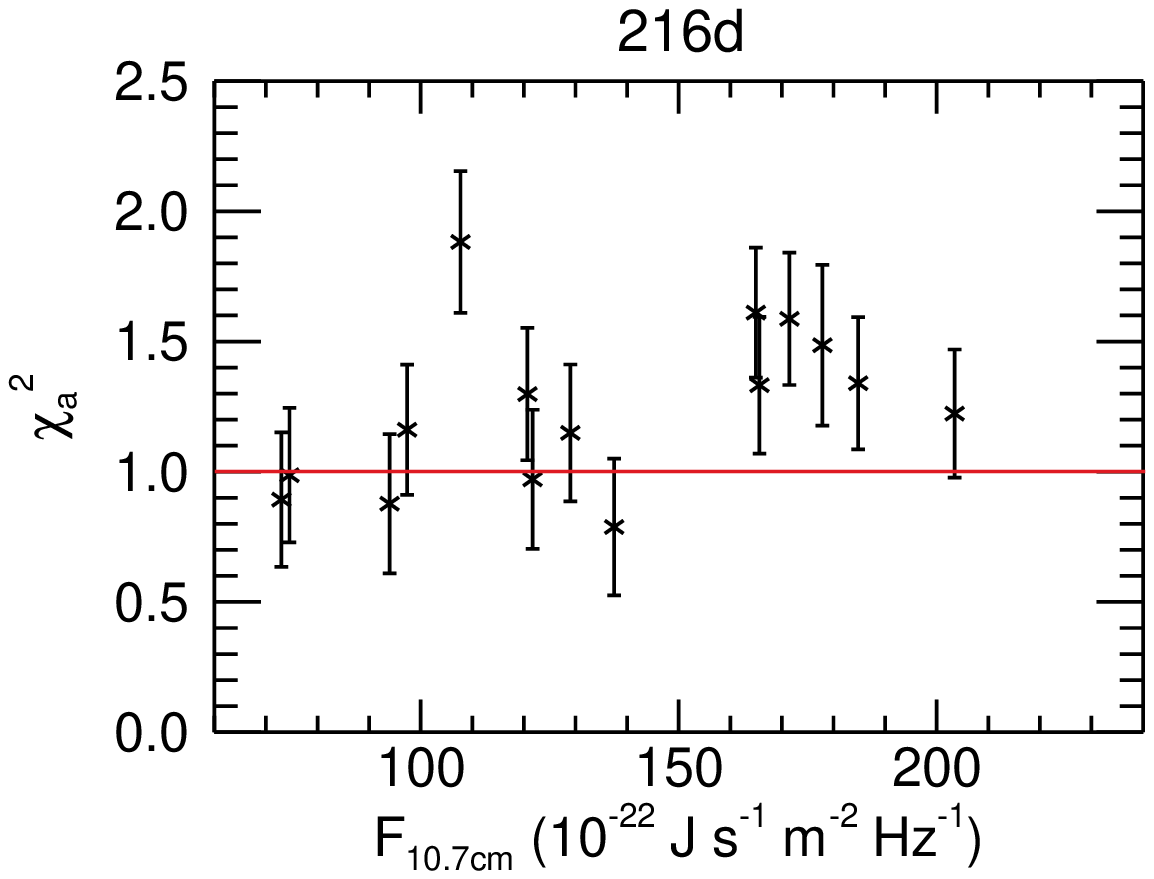}}
    {\includegraphics[clip,
    width=3.9cm]{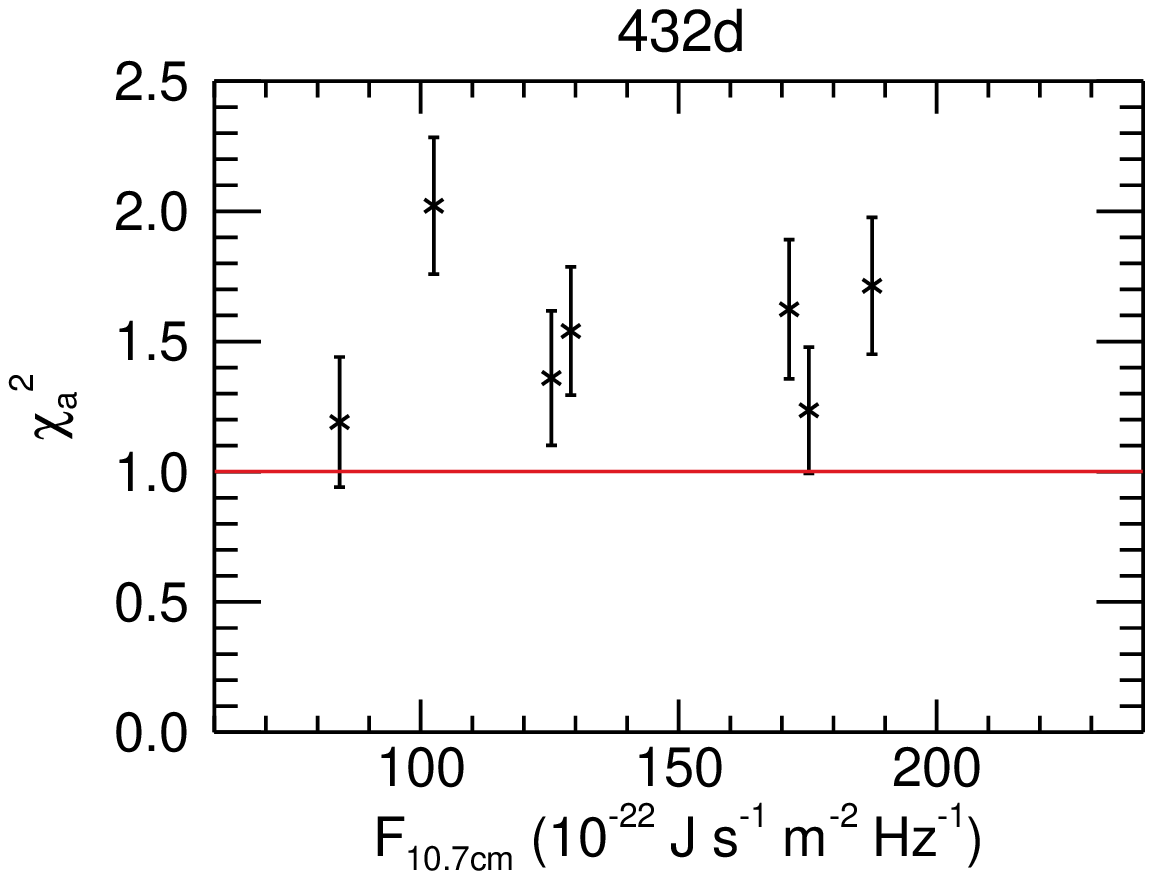}}
    \vspace{-0.3cm}
    %\subfigure[\scriptsize{$864\,\rm d$: Total reduced $\chi^2=1.41\pm0.07$}]
    %{\includegraphics[clip,
    %width=5.3cm, trim=-20mm 0mm -20mm 0mm]{chi_activity_f_864d}}\\
    \caption{The variation of $\chi_a^2$ with activity
    for time series of different lengths, when $108\,\rm d$ time series were used as the reference set.}
    \label{figure[chi a long]}
\end{figure}

We also investigated whether the accuracy of the predicted mode
frequencies was dependent on frequency (or $n$). Let $\chi_{n,l}^2$
be the reduced $\chi^2$ for individual $n$ and $l$ across all
activity levels. The results are shown in Figure \ref{figure[radial
chi]}. All of the proxies enable the frequencies to be predicted to
a reasonable degree of accuracy as $\chi_{n,l}^2$ is within
$3\sigma$ of unity for the vast majority of $n$ and $l$. Each panel
of Figure \ref{figure[radial chi]} contains an outlier, which is
observed because the frequency determined by the fitting procedure
is inaccurate since at certain times this mode contains little
power.

\begin{figure}
\centering
  \includegraphics[clip, width=3.9cm]{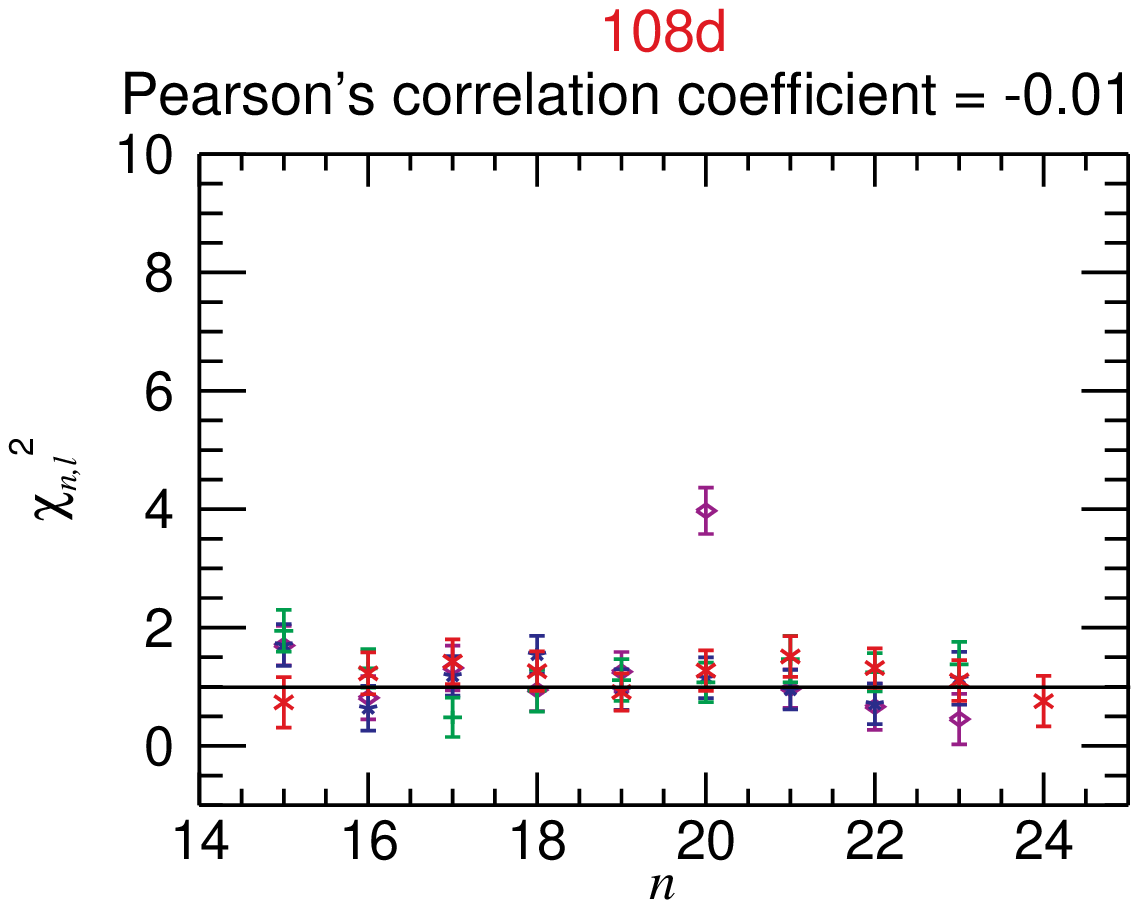}
  \includegraphics[clip, width=3.9cm]{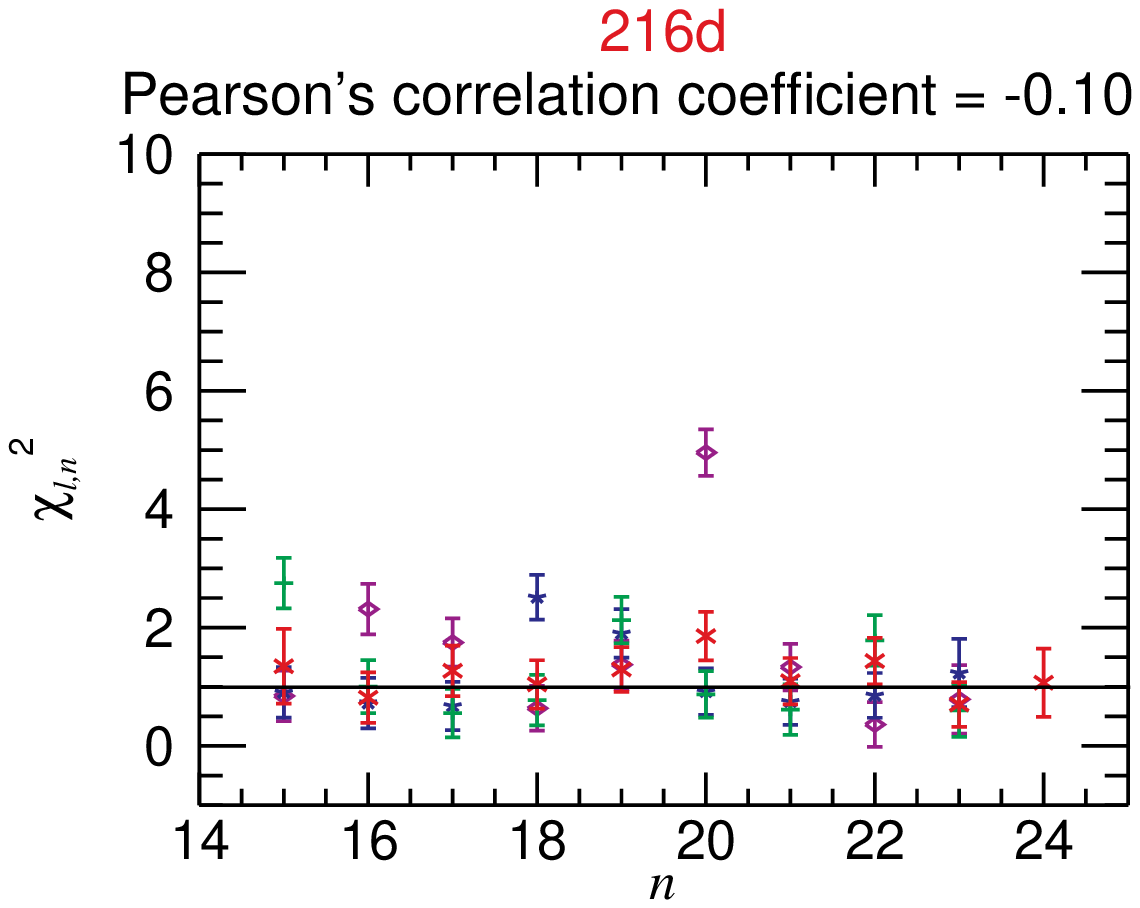}
  \includegraphics[clip, width=3.9cm]{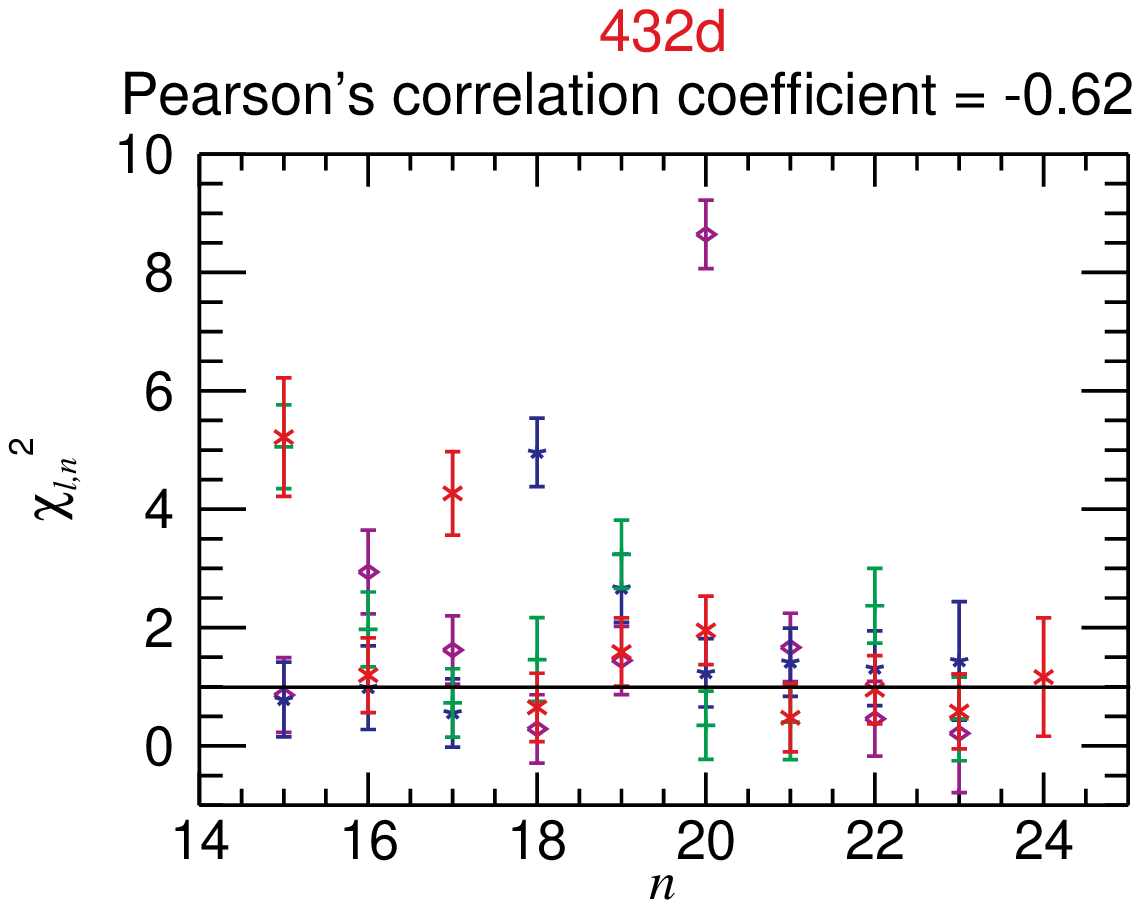}\\
    \includegraphics[clip, width=4.8cm]{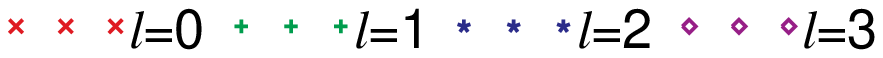}\\
\vspace{-0.3cm}
 \caption{The variation of $\chi_{n,l}^2$ with $n$
    for time series of different lengths, when $108\,\rm d$ time series were used as the reference set.}
    \label{figure[chi n long]}
\end{figure}

Although not shown here, the Pearson's correlation coefficients
between $\chi_{n,l}^2$ and $n$ were found to be approximately zero
implying the accuracy of the predictions is not dependent on $n$.
However, it should be noted that this is only true for the range of
$n$ examined here. Once we move outside this frequency range
$(2200\,\rm \mu Hz\le\nu\le3500\,\rm \mu Hz)$ the quality of both
the fitted frequencies and the predictions deteriorates
significantly.

\subsection{Time series of different lengths}

Next, we sought to determine whether the data observed in $108\,\rm
d$ time series in cycle 22 could be used to predict the acoustic
behaviour observed in longer ($216\,\rm d$, $432\,\rm d$) data sets
from cycle 23. Figure \ref{figure[chi a long]} shows the variation
of $\chi_a^2$ with activity when predictions were made for different
length time series. The determined $\chi_a^2$ increases with the
length of the time series, implying that the predictions for longer
time series are less accurate. Additionally, the weighted mean
$\chi_a^2$, which are quoted in Table \ref{table[longer for longer]}
(when the $108\,\rm d$ time series are used to make the
predictions), are found to be larger for the longer time series,
indicating the $108\,\rm d$ time series can not be used to predict
the frequencies observed in longer time series.

Figure \ref{figure[chi n long]} shows the variation of
$\chi_{l,n}^2$ with radial order, $n$, for different length time
series. The determined $\chi_{l,n}^2$ increase as $n$ decreases,
especially in the longer time series, indicating that the
predictions for low-$n$ modes are less accurate in long time series.
Quoted above each panel in Figure \ref{figure[chi n long]} are the
Pearson's correlation coefficients, which become increasingly
negative for the long time series and is in fact significant at a
95\% confidence level for the $432\,\rm d$ time series. This is
probably because it is more difficult to observe low-frequency modes
in $108\,\rm d$ time series than in longer data sets and so the
predicted frequencies are biased.

Finally we ask whether more accurate predictions are made if the
frequencies observed in longer time series are used to define the
relationship between frequency shift and activity. For example, we
have used the mode frequencies observed in time series of length
$216\,\rm d$ during cycle 22 to make predictions of the mode
frequencies observed in $216\,\rm d$ time series in cycle 23. The
observed weighted mean $\chi_a^2$, quoted in Table \ref{table[longer
for longer]}, indicate that the most accurate predictions are made
when the time series used to make the predictions and the time
series for which the predictions are required are both of the same
length.

\begin{table}[t]
   \centering
   \caption{The weighted mean $\chi_a^2$ between the observed and
   predicted frequency shifts for different length time series.}
   \vspace{0.1cm}
    \begin{tabular}{ccc}
     \hline
   \textbf{Time series length} & \textbf{Weighted mean $\chi_a^2$} & \textbf{Distance from $\chi_a^2=1$}\\
    \hline
    $108\,\rm d$ to predict $108\,\rm d$ & $1.18\pm0.06$ & $3.1\sigma$\\
    \hline
   $108\,\rm d$ to predict $216\,\rm d$ & $1.24\pm0.07$ & $3.5\sigma$ \\
    $216\,\rm d$ to predict $216\,\rm d$ & $1.06\pm0.07$ & $0.9\sigma$ \\
    \hline
    $108\,\rm d$ to predict $432\,\rm d$ & $1.51\pm0.10$ & $5.3\sigma$\\
    $216\,\rm d$ to predict $432\,\rm d$ & $1.13\pm0.10$ & $1.4\sigma$ \\
    $432\,\rm d$ to predict $432\,\rm d$ & $1.07\pm0.10$ & $0.8\sigma$ \\
    \hline
    \end{tabular}
    \label{table[longer for longer]}
\end{table}

\section{Discussion}\label{section[discussion]}

We have shown that, when $108\,\rm d$ time series are used, there
may be some evidence for a bias in the predictions, as the weighted
mean $\chi_a^2$ are noticeably greater than unity, indicating that
it may be problematic to use the $108\,\rm d$ time series to predict
the mode frequencies. The accuracy of the predictions made using the
$10.7\,\rm cm$ flux show least dependence on activity, probably
because the $10.7\,\rm cm$ flux is sensitive to both the strong and
the weak magnetic field. The predictions made using the KPMI were
not as accurate as those made using the other models and so this
proxy should not be used as a predictive tool.

The accuracy of the predictions decreases when $108\,\rm d$ time
series are used to predict mode frequencies for longer time series
and so $108\,\rm d$ time series can not be used to predict the
acoustic behaviour observed in longer time series. The predicted
frequencies were more accurate when longer data sets were used to
make the predictions and when the time series from cycle 22, which
were used to make the predictions, and the cycle 23 time series were
of the same length. Note that this analysis has been performed using
the mode frequencies observed in 2 solar cycles only but would
ideally use the data from more solar cycles.

\acknowledgements We thank past and present members of the BiSON
team and all those involved in the data collection process. The
authors acknowledge the financial support of the Science and
Technology Facilities Council (STFC).

\bibliographystyle{mn2e}
\bibliography{solar_cycle_ref}
%\begin{thebibliography}{}
%\bibitem[Chaplin et al. (1999)]{Chaplin1999}Chaplin, W.~J., Elsworth, Y., Isaak, G.~R., Miller, B.~A. and
%    New, R. 1999, MNRAS~308, 424
%\bibitem[Chaplin et al. (2004)]{Chaplin2004}Chaplin, W.~J., Elsworth, Y., Isaak, G.~R., Miller, B.~A. and
%    New, R. 2004, MNRAS~352, 1102
%\bibitem[Christensen-Dalsgaard \& Berthomieu (1991)\textbf{}]{Christensen1991}Christensen-Dalsgaard, J. and Berthomieu,
%G. 2001, in Theory of solar oscillations, 401
%\bibitem[de Toma et al. 2004]{Toma2004}de Toma, G., White, O.~R., Chapman, G.~A., Walton, S.~R.,
%Preminger, D.~G. and Cookson, A.~M. 2004, ApJ~609, 1140
%\bibitem[Woodard and Noyes, 1985]{Woodard1985}Woodard, M.~F. and Noyes, R.~W. 1985, Nature~318,
%449
%\end{thebibliography}

\end{document}